\documentclass[preprint,showpacs,preprintnumbers,amsmath,amssymb]{revtex4}

\usepackage{graphicx}
\usepackage{dcolumn}
\usepackage{bm}

\begin{document}

\title{The LISA Optimal Sensitivity}

\author{Thomas A. Prince}
\email{prince@srl.caltech.edu}
\altaffiliation [Also at: ]{Jet Propulsion Laboratory, California
  Institute of Technology, Pasadena, CA 91109}
\affiliation{Space Radiation Laboratory, California Institute of Technology,
Pasadena, CA 91125}

\author{Massimo Tinto}
\email{Massimo.Tinto@jpl.nasa.gov}
\altaffiliation [Also at: ]{Jet Propulsion Laboratory, California
  Institute of Technology, Pasadena, CA 91109}
\affiliation{Space Radiation Laboratory, California Institute of Technology,
Pasadena, CA 91125}

\author{Shane L. Larson}
\email{shane@srl.caltech.edu}
\affiliation{Space Radiation Laboratory, California Institute of Technology,
Pasadena, CA 91125}

\author{J.W. Armstrong}
\email{John.W.Armstrong@jpl.nasa.gov}
\affiliation{Jet Propulsion Laboratory, California Institute of Technology,
 Pasadena, CA 91109}

\date{\today}

\begin{abstract}
  The multiple Doppler readouts available on the Laser Interferometer
  Space Antenna (LISA) permit simultaneous formation of several
  interferometric observables.  All these observables are independent
  of laser frequency fluctuations and have different couplings to
  gravitational waves and to the various LISA instrumental noises.
  Within the functional space of interferometric combinations LISA
  will be able to synthesize, we have identified a triplet of
  interferometric combinations that show optimally combined
  sensitivity.  As an application of the method, we computed the
  sensitivity improvement for sinusoidal sources in the nominal,
  equal-arm LISA configuration.  In the part of the Fourier band where
  the period of the wave is longer than the typical light travel-time
  across LISA, the sensitivity gain over a single Michelson
  interferometer is equal to $\sqrt{2}$.  In the mid-band region,
  where the LISA Michelson combination has its best sensitivity, the
  improvement over the Michelson sensitivity is slightly better than
  $\sqrt{2}$, and the frequency band of best sensitivity is broadened.
  For frequencies greater than the reciprocal of the light
  travel-time, the sensitivity improvement is oscillatory and $\sim
  \sqrt{3}$, but can be greater than $\sqrt{3}$ near frequencies that
  are integer multiples of the inverse of the one-way light
  travel-time in the LISA arm.
\end{abstract}


\maketitle

\section{Introduction}

The Laser Interferometer Space Antenna (LISA) is a deep-space mission,
jointly proposed to the National Aeronautics and Space Administration
(NASA) and the European Space Agency (ESA), to detect and study
gravitational radiation in the millihertz frequency band.

LISA will use coherent laser beams exchanged between three remote,
widely separated, spacecraft. Modeling each spacecraft as carrying
lasers, beam splitters, photo-detectors, and drag-free proof masses on
each of two optical benches, it has been shown \cite{AET}, \cite{ETA},
\cite{TEA} that the six measured time series of Doppler shifts of the
one-way laser beams between spacecraft pairs, and the six measured
shifts between adjacent optical benches on each spacecraft, can be
combined, with suitable time delays, to cancel the otherwise
overwhelming frequency fluctuations of the lasers ($\Delta \nu/\nu
\simeq 10^{-13}/ \sqrt{\rm Hz}$), and the noise due to the mechanical
vibrations of the optical benches (which could be as large as $\Delta
\nu/\nu \simeq 10^{-16}/ \sqrt{\rm Hz}$). The achievable strain
sensitivity level $h \simeq 10^{-21} / \sqrt{\rm Hz}$ is set by the
buffeting of the drag-free proof masses inside each optical bench, and
by the shot noise at the photo-detectors.

In contrast to Earth-based, equi-arm interferometers for gravitational
wave detection, LISA will have multiple readouts. The data they
generate, when properly time shifted and linearly combined, provide
observables that are not only insensitive to laser frequency
fluctuations and optical bench motions, but also show different
couplings to gravitational radiation and to the remaining system
noises \cite{AET}, \cite{ETA}, \cite{TAE}. The space of all possible
interferometric combinations can be generated by properly combining
four generators \cite{AET}, and it has been shown to be an {\it
  algebraic module} \cite{DNV}. In this paper we derive from first
principles a particular combination of these generators which, for a
given waveform and source location in the sky, give maximum
signal-to-noise ratio.  In this respect, LISA should no longer be
regarded as a single detector, but rather as an array of
interferometers working in coincidence.

An outline of the paper is presented here. Section II provides a brief
summary of Time-Delay Interferometry, the data processing technique
needed to remove the frequency fluctuations of the six lasers used by
LISA.  After showing that the entire set of interferometric
combinations can be derived by properly combining four generators,
($\alpha, \beta, \gamma, \zeta$), in Section III we turn to the
problem of optimization of the Signal-to-Noise ratio (SNR) within this
functional space. As an application, we apply our results to the case
of sinusoidal signals randomly polarized and randomly distributed on
the celestial sphere. We find that the standard LISA sensitivity
figure derived for a single Michelson Interferometer \cite{ETA} can be
improved by a factor of $\sqrt{2}$ in the low-part of the frequency
band, and by more than $\sqrt{3}$ in the remaining part of the
accessible band.  In Section IV we present our comments and
conclusions.

\section{Time-Delay Interferometry for LISA}

Figure 1 shows the overall LISA geometry.  The spacecraft are labeled
$1$, $2$, $3$ and distances between pairs of spacecraft are $L_1$,
$L_2$, $L_3$, with $L_i$ being opposite spacecraft $i$.  Unit vectors
between spacecraft are $\hat n_i$, oriented as indicated in Figure 1.
We similarly index the relative frequency fluctuations $y_{ij}$ data
to be analyzed: $y_{31}$ is the relative frequency fluctuations time
series measured at reception at spacecraft $1$ with transmission from
spacecraft $2$ (along $L_3$).  Similarly, $y_{21}$ is the corresponding
time series derived from reception at spacecraft $1$ with transmission
from spacecraft $3$. The other four one-way relative frequency time
series from signals exchanged between the spacecraft are obtained by
cyclic permutation of the indices: $1$ $\to$ $2$ $\to$ $3$ $\to$ $1$.
The useful notation for delayed data streams will be also used:
$y_{31,2} = y_{31}(t - L_2)$, $y_{31,23} = y_{31}(t - L_2 - L_3) =
y_{31,32}$, etc. (units in which $c = 1$).  Six more Doppler difference
series result from laser beams exchanged between adjacent optical
benches within each spacecraft; these are similarly indexed as
$z_{ij}$ ($i,j = 1, 2, 3 \ ; \ i \ne j$) (see \cite{ETA} and
\cite{TEA} for details).

\begin{figure}
\centering
\includegraphics[width=2.5in, angle=-90]{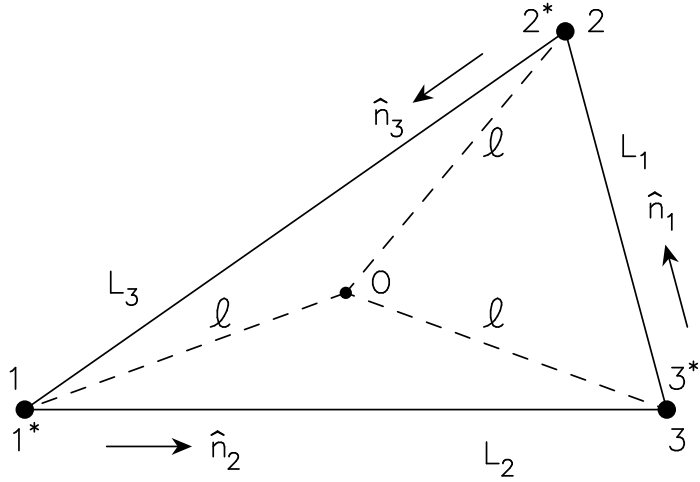}
\caption{Schematic LISA configuration.  Each spacecraft is equidistant
from the point O, in the plane of the spacecraft.  Unit vectors
$\hat n_i$ point between spacecraft pairs with the indicated
orientation.  At each vertex spacecraft there are two optical
benches (denoted 1, $1^*$, etc.), as indicated.}
\end{figure}

The light paths for the $y_{i1}$'s and $z_{i1}$'s can be traced in
Figure 2.  An outgoing light beam transmitted to a distant spacecraft
is routed from the laser on the local optical bench using mirrors and
beam splitters; this beam does not interact with the local proof mass.
Conversely, an {\it {incoming}} light beam from a distant spacecraft
is bounced off the local proof mass before being reflected onto the
photo-detector where it is mixed with light from the laser on that
same optical bench.  These data are denoted $y_{31}$ and $y_{21}$ in
Figure 2.  Beams exchanged between adjacent optical benches however do
precisely the opposite.  Light to be {\it {transmitted}} from the
laser on an optical bench is {\it {first}} bounced off the proof mass
it encloses and then directed via fiber optics to the other optical
bench.  Upon reception it does {\it not} interact with the proof mass
there, but is directly mixed with local laser light.  They are
$z_{31}$ and $z_{21}$ in Figure 2.

The frequency fluctuations introduced by the gravitational wave
signal, the lasers, the optical benches, the proof masses, the fiber
optics, and the measurement itself at the photo-detector (shot noise)
enter into the Doppler observables $y_{ij}$, $z_{ij}$ with specific
time signatures. They have been derived in the literature \cite{ETA},
\cite{TEA}, and we refer the reader to those papers for a detailed
discussion.  The Doppler data $y_{ij}$, $z_{ij}$ are the fundamental
measurements needed to synthesize all the interferometric observables
unaffected by laser and optical bench noises.

\begin{figure}
\centering
\includegraphics[width=5in, angle=0]{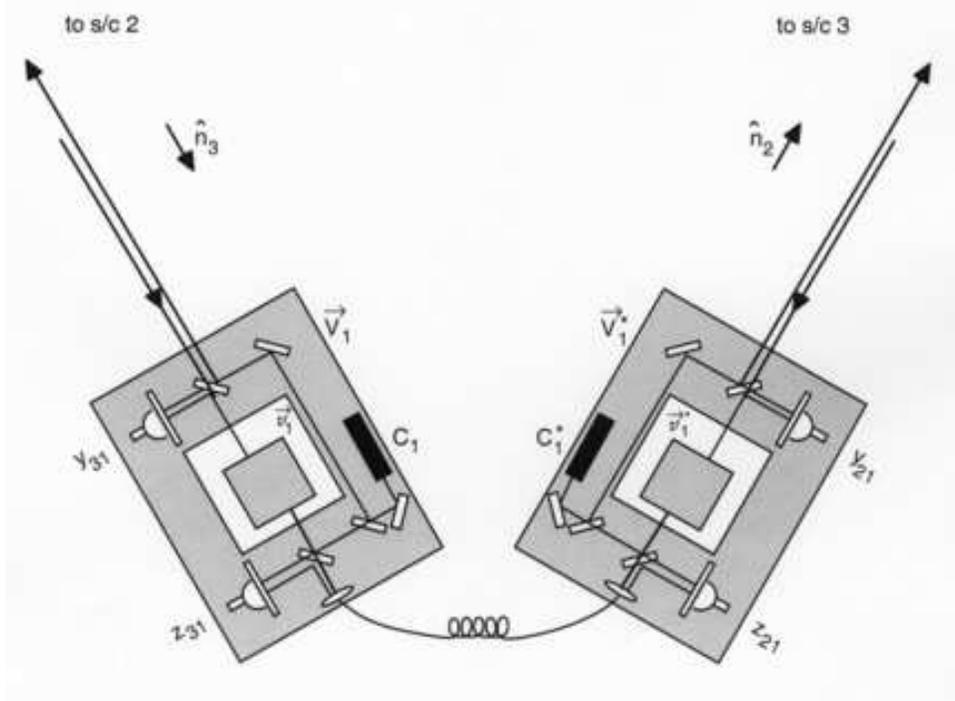}
\caption{Schematic diagram of proof-masses-plus-optical-benches
  for a LISA spacecraft.  The left-hand bench reads out the Doppler
  signals $y_{31}$ and $z_{31}$.  The right hand bench analogously
  reads out $y_{21}$ and $z_{21}$.  The random velocities of the two
  proof masses and two optical benches are indicated (lower case $\vec
  v_i$ for the proof masses, upper case $\vec V_i$ for the optical
  benches.)}
\end{figure}
The simplest such combination, the totally symmetrized Sagnac
response $\zeta$, uses all the data of Figure 2 symmetrically
\begin{eqnarray}
\zeta & = & y_{32,2} - y_{23,3} + y_{13,3} - y_{31,1} + y_{21,1} - y_{12,2}
\nonumber \\
& & +
{1 \over 2} (-{z_{13,21}} + {z_{23,12}} - 
{z_{21,23}} + {z_{31,23}} - {z_{32,13}} + {z_{12,13}})
\nonumber \\
& & +
{1 \over 2} (-{z_{32,2}} + {z_{12,2}} - {z_{13,3}} + 
{z_{23,3}} - {z_{21,1}} + {z_{31,1}}) \ ,
\label{eq:1}
\end{eqnarray}
and its transfer functions to gravitational waves and instrumental
noises were derived in \cite{AET}, and \cite{ETA} respectively.  In
particular, $\zeta$ has a ``six-pulse response'' to gravitational
radiation, i.e.  a $\delta$-function gravitational wave signal
produces six distinct pulses in $\zeta$ \cite{AET}, which are located
with relative times depending on the arrival direction of the wave and
the detector configuration.

Together with $\zeta$, three more interferometric combinations,
($\alpha, \beta, \gamma$), jointly generate the entire space of
interferometric combinations \cite{AET}, \cite{ETA}, \cite{DNV}. Their
expressions in terms of the measurements $y_{ij}$, $z_{ij}$ are as
follows
\begin{eqnarray}
\alpha & = & y_{21} - y_{31} + y_{13,2} - y_{12,3} + y_{32,12} - y_{23,13}
\nonumber \\
& & -
{1 \over 2} ({z_{13,2}} + {z_{13,13}} + {z_{21}} + 
{z_{21,123}} + {z_{32,3}} + {z_{32,12}})
\nonumber \\
& & +
{1 \over 2} ({z_{23,2}} + {z_{23,13}} + {z_{31}} + 
{z_{31,123}} + {z_{12,3}} + {z_{12,12}}) \ ,
\label{eq:2}
\end{eqnarray}
\noindent
with $\beta$, and $\gamma$ derived by permuting the spacecraft
indices in $\alpha$. Like in the case of $\zeta$, a $\delta$-function
gravitational wave produces six pulses in $\alpha$, $\beta$, and
$\gamma$.

We should remind the reader that the four interferometric
responses ($\alpha, \beta, \gamma, \zeta$) satisfy the following
relationship
\begin{equation}
\zeta - \zeta_{,123} =  \alpha_{,1} - \alpha_{,23} + \beta_{,2} -
\beta_{,13} + \gamma_{,3} - \gamma_{,12} \ .
\label{eq:3}
\end{equation}
Jointly they also give the expressions of the interferometric
combinations derived in \cite{AET}, \cite{ETA}: the Unequal-arm
Michelson (${\rm X}, {\rm Y}, {\rm Z}$), the Beacon (${\rm P}, {\rm
  Q}, {\rm R}$), the Monitor (${\rm E}, {\rm F}, {\rm G}$), and the
Relay (${\rm U}, {\rm V}, {\rm W}$) responses
\begin{eqnarray}
{\rm X}_{,1} & = & \alpha_{,23} - \beta_{,2} - \gamma_{,3} + \zeta \ ,
\label{eq:4} \\
{\rm P} & = & \zeta - \alpha_{,1} \ ,
\label{eq:5} \\
{\rm E} & = & \alpha - \zeta_{,1} \ ,
\label{eq:6} \\
{\rm U} & = & \gamma_{,1} - \beta \ ,
\label{eq:7}
\end{eqnarray}
with the remaining expressions obtained from equations (\ref{eq:4},
\ref{eq:5}, \ref{eq:6}, \ref{eq:7}) by permutation of the spacecraft
indices. All these interferometric combinations have been shown to add
robustness to the mission with respect to failures of subsystems, and
potential design, implementation, or cost advantages \cite{AET},
\cite{ETA}.

\section{Optimal Sensitivity for LISA}

All the above interferometric combinations have been shown to
individually have rather different sensitivities \cite{ETA}, as a
consequence of their different responses to gravitational radiation
and system noises. LISA has the capability of {\underbar
  {simultaneously}} observing a gravitational wave signal with many
different interferometric combinations, all having different antenna
patterns and noises. We should thus no longer regard LISA as a single
detector system but rather as an array of gravitational wave detectors
working in coincidence. This suggests that the presently adopted LISA
sensitivity could be improved by {\it optimally} combining the four
generators ($\alpha, \beta, \gamma, \zeta$). In mathematical terms
this can be restated by saying that we should be able to find that
particular combination of the four generators that has maximum
signal-to-noise ratio to a given gravitational wave signal.

In order to proceed with this idea, let us consider the following
linear combination $\eta (f)$ of the Fourier transforms of the four
generators (${\widetilde {\alpha}}, {\widetilde {\beta}}, {\widetilde
  {\gamma}}, {\widetilde {\zeta}}$)
\begin{equation}
\eta(f) \equiv a_1 (f, {\vec \lambda}) \ {\widetilde{\alpha}} (f) \ +
\ a_2 (f, {\vec \lambda}) \ {\widetilde{\beta}} (f) \ +
\ a_3 (f, {\vec \lambda}) \ {\widetilde{\gamma}} (f) \ +
\ a_4 (f, {\vec \lambda}) \ {\widetilde{\zeta}} (f)  \ ,
\label{eq:8}
\end{equation}
where the $\{a_i (f, \vec \lambda)\}^4_{i=1}$ are arbitrary complex
functions of the Fourier frequency $f$, and of a vector $\vec \lambda$
containing parameters characterizing the gravitational wave signal
(source location in the sky, waveform parameters, etc.) and the noises
affecting the four responses (noise levels, their correlations, etc.).
For a given choice of the four functions $\{a_i \}^4_{i=1}$, $\eta$
gives an element of the functional space of interferometric
combinations generated by ($\alpha, \beta, \gamma, \zeta$). Our goal
is therefore to identify, for a given gravitational wave signal, the
four functions $\{a_i \}^4_{i=1}$ that maximize the signal-to-noise
ratio, $SNR_{\eta}^2$, of the combination $\eta$
\begin{equation}
SNR_{\eta}^2 = 
\int_{f_{l}}^{f_u} 
{
{|a_1 \ {\widetilde \alpha_s} + a_2 \ {\widetilde \beta_s} + a_3 \
  {\widetilde \gamma_s} + a_4 {\widetilde \zeta_s} |^2} 
\over
{\langle|a_1 \ {\widetilde \alpha_n} + a_2 \ {\widetilde \beta_n} +
  a_3 \ {\widetilde \gamma_n} + a_4 {\widetilde \zeta_n} |^2 \rangle} } \ df \ .
\label{eq:9bis}
\end{equation}
In equation (\ref{eq:9bis}) the subscripts $s$ and $n$ refer to the
signal and the noise parts of (${\widetilde {\alpha}}, {\widetilde
  {\beta}}, {\widetilde {\gamma}}, {\widetilde {\zeta}}$)
respectively, the angle brackets represent noise ensemble averages,
and the interval of integration ($f_l, f_u$) corresponds to the
frequency band accessible by LISA.

Before proceeding with the maximization of the $SNR_{\eta}^2$ we may
notice from equation (\ref{eq:3}) that the Fourier transform of the
totally symmetric Sagnac combination, $\widetilde \zeta$, multiplied
by the transfer function $1 - e^{2 \pi i f (L_1 + L_2 + L_3)}$ can be
written as a linear combination of the Fourier transforms of the
remaining three generators ($ {\widetilde {\alpha}}, {\widetilde
  {\beta}}, {\widetilde {\gamma}}$). Since the signal-to-noise ratio
of $\eta$ and $(1 - e^{2 \pi i f (L_1 + L_2 + L_3)}) \eta$ are equal,
we may conclude that the optimization of the signal-to-noise ratio of
$\eta$ can be performed only on the three observables $\alpha, \beta,
\gamma$. This implies the following redefined expression for
$SNR_{\eta}^2$
\begin{equation}
SNR_{\eta}^2 = 
\int_{f_{l}}^{f_u} 
{
{|a_1 \ {\widetilde \alpha_s} + a_2 \ {\widetilde \beta_s} + a_3 \
  {\widetilde \gamma_s} |^2} 
\over
{\langle|a_1 \ {\widetilde \alpha_n} + a_2 \ {\widetilde \beta_n} +
  a_3 \ {\widetilde \gamma_n} |^2 \rangle} } \ df \ .
\label{eq:9}
\end{equation}
The $SNR_{\eta}^2$ can be regarded as a functional over the space of
the three complex functions $\{ a_i \}^3_{i=1}$, and the particular
set of complex functions that extremize it can of course be derived by
solving the associated set of Euler-Lagrange equations.

In order to make the derivation of the optimal SNR easier, let us
first denote by ${\bf x}^{(s)}$ and ${\bf x}^{(n)}$ the two vectors of
the signals (${\widetilde{\alpha}}_s, {\widetilde{\beta}}_s,
{\widetilde{\gamma}}_s$) and the noises (${\widetilde{\alpha}}_n,
{\widetilde{\beta}}_n, {\widetilde{\gamma}}_n$) respectively. Let us also
define $\bf a$ to be the vector of the three functions $\{a_i
\}^3_{i=1}$, and denote with ${\bf C}$ the hermitian, non-singular,
correlation matrix of the vector random process ${\bf x}_n$
\begin{equation}
({\bf C})_{rt} \equiv \langle {\bf x}^{(n)}_{r} {\bf x}^{(n)*}_{t} \rangle \ .
\label{eq:14}
\end{equation}
If we finally define $({\bf A})_{ij}$ to be the components of the
hermitian matrix ${\bf x}^{(s)}_i {\bf x}^{(s)*}_j$, we can rewrite
$SNR_{\eta}^2$ in the following form
\begin{equation}
SNR_{\eta}^2 = 
\int_{f_{l}}^{f_u} 
{
{{\bf a}_i {\bf A}_{ij} {\bf a}^*_j }
\over
{{\bf a}_r {\bf C}_{rt} {\bf a}^*_t }
} \ df \ ,
\label{eq:16}
\end{equation}
where we have adopted the usual convention of summation over repeated
indices. Since the noise correlation matrix ${\bf C}$ is non-singular,
and the integrand is positive definite or null, the stationary values
of the signal-to-noise ratio will be attained at the stationary values
of the integrand, which are given by solving the following set of
equations (and their complex conjugated expressions)
\begin{equation}
{\partial \over {\partial {\bf a}_k}} \ \left[
{
{{\bf a}_i {\bf A}_{ij} {\bf a}^*_j }
\over
{{\bf a}_r {\bf C}_{rt} {\bf a}^*_t }
} \right] = 0 \ \ \ , \ \ \ k = 1, 2, 3 \ .
\label{eq:16bis}
\end{equation}
After taking the partial derivatives, equation (\ref{eq:16bis}) can be
rewritten in the following form
\begin{equation}
({\bf C}^{-1})_{ir} 
({\bf A})_{rj} ({\bf a}^*)_j = \left[
{
{{\bf a}_p {\bf A}_{pq} {\bf a}^*_q }
\over
{{\bf a}_l {\bf C}_{lm} {\bf a}^*_m }
} \right]  
\ ({\bf a}^*)_i \ \ \ , \ \ \ i = 1, 2, 3 
\label{eq:16tris}
\end{equation}
which tells us that the stationary values of the signal-to-noise ratio
of $\eta$ are equal to the eigenvalues of the the matrix ${\bf C^{-1}
  \cdot A}$. The result in equation (\ref{eq:16bis}) is well known in
the theory of quadratic forms, and it is called the Rayleigh's
principle \cite{Noble69}, \cite{Selby64}.

In order now to identify the eigenvalues of the matrix ${\bf C^{-1}
  \cdot A}$, we first notice that the $3 \times 3$ matrix ${\bf A}$
has rank $1$. This implies that the matrix ${\bf
  C}^{-1} \cdot {\bf A}$ has also rank $1$, as it is easy to verify.
Therefore two of its three eigenvalues are equal to zero, while the
remaining non-zero eigenvalue represents the solution we are looking
for.

The analytic expression of the third eigenvalue can be obtained by
using the property that the trace of the $3 \times 3$ matrix ${\bf
  C}^{-1} \cdot {\bf A}$ is equal to the sum of its three eigenvalues,
and in our case to the eigenvalue we are looking for.  From these
considerations we derive the following expression for the optimized
signal-to-noise ratio ${SNR_{\eta}^2}_{\rm opt.}$
\begin{equation}
{SNR_{\eta}^2}_{\rm opt.} = \int_{f_{l}}^{f_u} 
{\bf x}^{(s)*}_i \ ({\bf C}^{-1})_{ij} \ {\bf x}^{(s)}_j  \ df \ .
\label{eq:18}
\end{equation}
\noindent
We can summarize the results derived in this section, which are given
by equations (\ref{eq:9},\ref{eq:18}), in the following way:

\noindent
(i) among all possible interferometric combinations LISA will be able
to synthesize with its four generators $\alpha, \beta, \gamma, \zeta$,
the particular combination giving maximum signal-to-noise ratio can be
obtained by using only three of them, namely ($\alpha, \beta,
\gamma$);

\noindent
(ii) the expression of the optimal signal-to-noise ratio given by
equation (\ref{eq:18}) implies that LISA should be regarded as a
network of three interferometer detectors of gravitational radiation
(of responses ($\alpha, \beta, \gamma$)) working in coincidence
\cite{Finn01}.

\subsection{Application}

As an application of equation (\ref{eq:18}), here we calculate the
sensitivity that LISA can reach when observing sinusoidal signals
uniformly distributed on the celestial sphere and of random
polarization. In order to calculate the optimal signal-to-noise ratio
we will also need to use a specific expression for the noise
correlation matrix ${\bf C}$. As a simplification, we will assume the
LISA arm-lengths to be equal to its nominal value $L = 16.67 \ {\rm
  sec.}$, the optical-path noises to be equal and uncorrelated to each
other, and finally the noises due to the proof-mass noises to be also
equal, uncorrelated to each other and to the optical-path noises.
Under these assumptions the correlation matrix becomes real, its three
diagonal elements are equal, and all the off-diagonal terms are equal
to each other, as it is easy to verify by direct calculation
\cite{ETA}. The noise correlation matrix ${\bf C}$ is therefore
uniquely identified by two real functions, $S_{\alpha}$ and $S_{\alpha
  \beta}$, in the following way
\[ {\bf C} = \left( \begin{array}{ccc}
S_{\alpha} & S_{\alpha \beta} & S_{\alpha \beta} \\
S_{\alpha \beta} & S_{\alpha} & S_{\alpha \beta} \\
S_{\alpha \beta} & S_{\alpha \beta} & S_{\alpha} 
\end{array} 
\label{eq:19}
\right).\] 

The expression of the optimal signal-to-noise ratio assumes a rather
simple form if we diagonalize this correlation matrix by properly
``choosing a new basis''. There exists an orthogonal
transformation of the generators ($ {\widetilde {\alpha}}, {\widetilde
  {\beta}}, {\widetilde {\gamma}}$) which will transform the optimal
signal-to-noise ratio into the sum of the signal-to-noise ratios of
the ``transformed'' three interferometric combinations.  The expressions
of the three eigenvalues $\{\mu_i\}^3_{i=1}$ (which are real) of the
noise correlation matrix ${\bf C}$ can easily be found by using the
algebraic manipulator {\it Mathematica} \cite{Wolf02}, and they are
equal to
\begin{equation}
\mu_1 = \mu_2 = S_{\alpha} - S_{\alpha \beta} \ \ \ , \ \ \ 
\mu_3 = S_{\alpha} + 2 \ S_{\alpha \beta} \ .
\label{eq:20}
\end{equation}
Note that two of the three real eigenvalues, ($\mu_1, \mu_2$), are
equal. This implies that the eigenvector associated to $\mu_3$ is
orthogonal to the two-dimensional space generated by the eigenvalue
$\mu_1$, while any chosen pair of eigenvectors corresponding to
$\mu_1$ will not necessarily be orthogonal.  This inconvenience can
be avoided by choosing an arbitrary set of vectors in this
two-dimensional space, and by ortho-normalizing them. After some simple
algebra, we have derived the following three ortho-normalized
eigenvectors
\begin{equation}
{\bf v_1} = {1 \over {\sqrt{2}}} \ (-1, 0, 1) \ \ \ , \ \ \ 
{\bf v_2} = {1 \over {\sqrt{6}}} \ (1, -2, 1) \ \ \ , \ \ \ 
{\bf v_3} = {1 \over {\sqrt{3}}} \ (1, 1, 1) \ .
\label{eq:21}
\end{equation}
Equation (\ref{eq:21}) implies the following three linear combinations of the
generators (${\widetilde{\alpha}}, {\widetilde{\beta}},
{\widetilde{\gamma}}$)
\begin{equation}
A = {1 \over \sqrt{2}} \ ({\widetilde{\gamma}} - {\widetilde{\alpha}}) \ \ \ , \ \ \ 
E = {1 \over \sqrt{6}} \ ({\widetilde{\alpha}} - 2 {\widetilde{\beta}} + {\widetilde{\gamma}}) \ \ \ , \ \ \ 
T = {1 \over \sqrt{3}} \ ({\widetilde{\alpha}} + {\widetilde{\beta}} +
{\widetilde{\gamma}}) \ ,
\label{eq:22}
\end{equation}
where $A$, $E$, and $T$ are italicized to indicate that these are
``orthogonal modes''. Although the expressions for the modes $A$ and
$E$ depend on our particular choice for the two eigenvectors (${\bf
  v_1}, {\bf v_2}$), it is clear from our earlier considerations that
the value of the optimal signal-to-noise ratio is unaffected by such a
choice. From equation (\ref{eq:22}) it is also easy to verify that the
noise correlation matrix of these three combinations is diagonal, and
that its non-zero elements are indeed equal to the eigenvalues given
in equation (\ref{eq:20}).

In order to calculate the sensitivity corresponding to the expression
of the optimal signal-to-noise ratio, we have proceeded similarly to
what was done in \cite{AET}, \cite{ETA}, and described in more detail
in \cite{WP02}. We assume an equal-arm LISA ($L = 16.67$ light
seconds), and take the one-sided spectra of proof mass and aggregate
optical-path-noises (on a single link), expressed as fractional
frequency fluctuation spectra, to be (\cite{ETA}, \cite{PPA98}),
$S^{proof\ mass}_y = 2.5 \times {10^{-48}} \ {\left[f/{1 Hz}
  \right]}^{-2} \ Hz^{-1}$ and $S^{optical \ path}_y = 1.8 \times
{10^{-37}} \ {\left[f/1 Hz \right]}^2 \ Hz^{-1}$, respectively.  We
also assume that aggregate optical path noise has the same transfer
function as shot noise.

The optimum SNR is the square root of the sum of the squares of the
SNRs of the three ``orthogonal modes'' ($A, E, T$).  To compare with previous
sensitivity curves of a single LISA Michelson interferometer, we
construct the SNRs as a function of Fourier frequency for sinusoidal
waves from sources uniformly distributed on the celestial sphere. To
produce the SNR of each of the ($A, E, T$) modes we need the
gravitational wave response and the noise response as a function of
Fourier frequency.  We build up the gravitational wave responses of
the three modes ($A, E, T$) from the gravitational wave responses of
($\alpha, \beta, \gamma$).  For $7000$ Fourier frequencies in the
${\sim}10^{-4}$ Hz to ${\sim}1$ Hz LISA band, we produce the Fourier
transforms of the gravitational wave response of ($\alpha, \beta,
\gamma$) from the formulas in \cite{AET}, \cite{WP02}.  The averaging
over source directions (uniformly distributed on the celestial sphere)
and polarization states (uniformly distributed on the Poincar\'e
sphere) is performed via a Monte Carlo method. From the Fourier
transforms of the ($\alpha, \beta, \gamma$) responses at each
frequency, we construct the Fourier transforms of ($A, E, T$).  We
then square and average to compute the mean-squared responses of
($A, E, T$) at that frequency from $10^4$ realizations of (source
position, polarization state) pairs.

The noise spectra of ($A, E, T$) are determined from the raw spectra of 
proof-mass and optical-path noises, and the transfer functions of these
noises to ($A, E, T$).  Using the transfer functions given in \cite{ETA},
the resulting spectra are equal to
\begin{eqnarray}
S_{A}(f) = S_{E}(f) & = & \ 16 \ \sin^2(\pi f L) \
[3 + 2 \cos(2 \pi f L) + \cos(4 \pi f L)] S^{proof\ mass}_y(f)
\nonumber \\
& + & \ 8 \ \sin^2(\pi f L) \ [2 + \cos(2 \pi f L)] \ S^{optical \
  path}_y(f) \ ,
\end{eqnarray}

\begin{eqnarray}
S_{T}(f) & = & \ 2 [1 + 2 \cos (2 \pi f L)]^2 \ 
[4 \ \sin^2 (\pi f L) S_y^{proof\ mass} + S_y^{optical\ path} (f)] \ .
\end{eqnarray}

Let the amplitude of the sinusoidal gravitational wave be $h$.  The
SNR for, e.g. $A$, $SNR_{A}$, at each frequency $f$ is equal to $h$
times the ratio of the root-mean-squared gravitational wave response
at that frequency divided by $\sqrt{S_{A}(f) \ B}$, where $B$ is the
bandwidth conventionally taken to be equal to $1$ cycle per year.
Finally, if we take the reciprocal of $SNR_{A}/h$ and multiply it by
$5$ to get the conventional $SNR = 5$ sensitivity criterion, we obtain
the sensitivity curve for this combination which can then be compared
against the corresponding sensitivity curve for the equal-arm
Michelson Interferometer.

In Figure 3 we show the sensitivity curve for the LISA equal-arm
Michelson response ($SNR = 5$) as a function of the Fourier frequency,
and the sensitivity curve from the optimum weighting of the data
described above: $5 h / \sqrt{SNR_{A}^2 + SNR_{E}^2 + SNR_{T}^2}$. The
SNRs were computed for a bandwidth of 1 cycle/year.  Note that at
frequencies where the LISA Michelson combination has best sensitivity,
the improvement in signal-to-noise ratio provided by the optimal
combination is slightly larger than $\sqrt{2}$.
\begin{figure}
\centering
\includegraphics[width=2.5in, angle=-90]{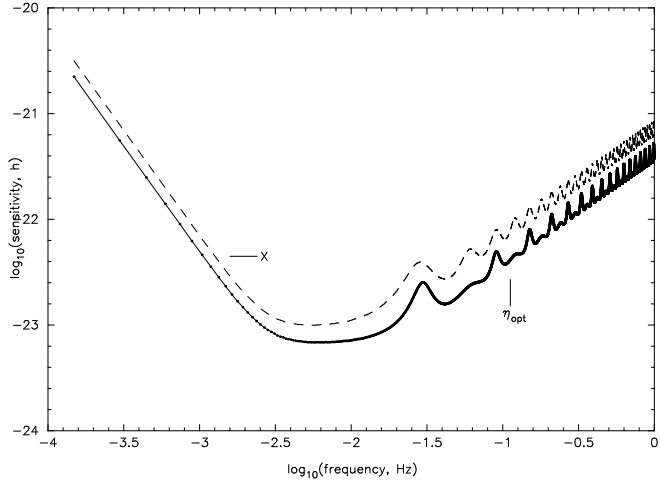}
\caption{The LISA Michelson sensitivity curve (SNR = 5)
and the sensitivity curve for the optimal combination of the data, both
as a function of Fourier frequency. The integration time is equal to
one year, and LISA is assumed to have a nominal armlength $L =
16.67 {\rm sec}$.}
\end{figure}

In Figure 4 we plot the ratio between the optimal SNR and the SNR of a
single Michelson interferometer.  In the long-wavelength limit, the
SNR improvement is $\sqrt{2}$.  For Fourier frequencies greater than
or about equal to $1/L$, the SNR improvement is larger and varies
with the frequency, showing an average value of about $\sqrt{3}$.  In
particular, for bands of frequencies centered on integer multiples of
$1/L$, $SNR_{T}$ contributes strongly and the aggregate SNR in these
bands can be greater than $2$.
\begin{figure}
\centering
\includegraphics[width=2.5in, angle=-90]{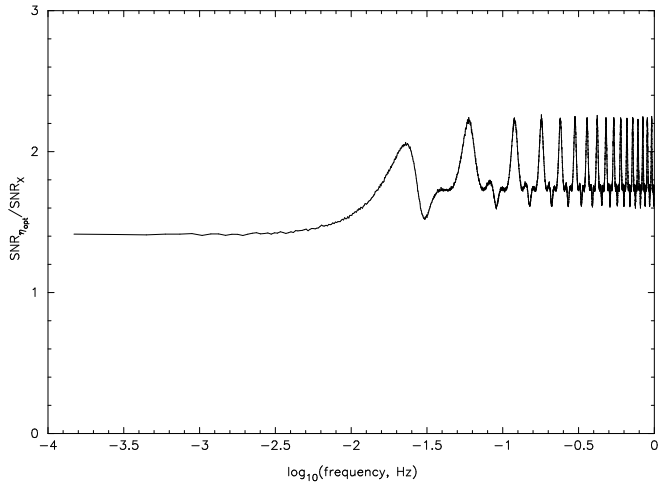}
\caption{The optimal SNR divided by the SNR of a single Michelson 
  interferometer, as a function of the Fourier frequency $f$. The
  sensitivity gain in the low-frequency band is equal to $\sqrt{2}$,
  while it can get larger than $2$ at selected frequencies in the
  high-frequency region of the accessible band. The integration time
  has been assumed to be one year, and the proof mass and optical path
  noise spectra are the nominal ones. See the main body of the paper
  for a quantitative discussion of this point.}
\end{figure}
In order to better understand the contribution from the three
different combinations to the optimal combination of the three
generators, in Figure 5 we plot the signal-to-noise ratios of ($A, E,
T$) as well as the optimal signal-to-noise ratio.  For an assumed $h =
10^{-23}$, the SNRs of the three modes are plotted versus frequency.
For the equal-arm case computed here, the SNRs of $A$ and $E$ are
equal across the band.  In the long wavelength region of the band,
modes $A$ and $E$ have SNRs much greater than mode $T$, where its
contribution to the total SNR is negligible. At higher frequencies,
however, the $T$ combination has SNR greater than or comparable to the
other modes and can dominate the SNR improvement at selected
frequencies.
\begin{figure}
\centering
\includegraphics[width=2.5in, angle=-90]{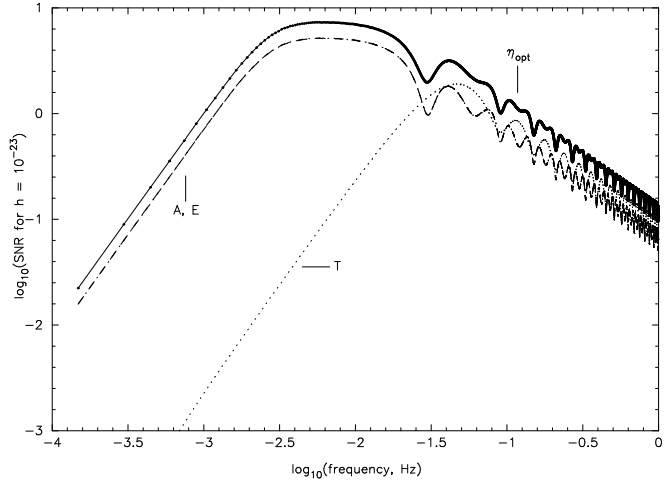}
\caption{The SNRs of the three combinations, ($A, E, T$), and their sum
  as a function of the Fourier frequency $f$. The SNRs of $A$ and $E$
  are equal over the entire frequency band. The SNR of $T$ is
  significantly smaller than the other two in the low part of the
  frequency band, while is comparable to (and at times larger than)
  the SNR of the other two in the high-frequency region. See text for
  a complete discussion.}
\end{figure}

\section{Conclusions}

The use of Time-Delay Interferometry has shown that LISA has the
capability to simultaneously observe gravitational waves in the
millihertz band with several, and rather different, interferometric
data combinations. In this paper we have identified, for a given (but
otherwise arbitrary) gravitational wave signal, the particular
interferometric combination that gives maximum signal-to-noise ratio.
In this context we have actually shown that LISA should no longer be
regarded as a single-instrument mission, but rather as a network of
interferometer detectors of gravitational radiation working in
coincidence.  We have identified the general expression of the optimal
combination of the generators ($\alpha, \beta, \gamma$), which should
be used when observing a specified gravitational wave signal.

Under rather general assumptions on the properties of the noise
correlation matrix, and for sinusoidal gravitational wave signals that
are randomly distributed over the celestial sphere and over the
polarization states, we have found that the sensitivity gain of the
optimal data combination over that of the Michelson combination can be
significant, and it varies over the frequency band accessible by LISA.
In the low part of the frequency band such improvement is equal to
$\sqrt{2}$, and it grows to values larger than $\sqrt{3}$ at higher
frequencies and in small frequency bands centered on frequencies that
are integer multiple of the inverse of the one-way light-travel time.

The results derived in this paper will have immediate application to
the solution of the so called ``Inverse Problem'' for LISA, that is to
say the determination of the source location and of the wave's two
independent amplitudes from the data LISA will be able to generate. We
will estimate the accuracies in the determination of the source
location and of the wave's amplitudes that the optimal combination of
the three interferometric responses ($\alpha, \beta, \gamma$) derived
in this paper will imply. This work is in progress, and will be
presented in a follow-up publication.

\begin{acknowledgments}
  We would like to thank Drs. Albert Lazzarini and Frank B. Estabrook
  for stimulating conversations while this work was in progress. The
  research was performed at the Jet Propulsion Laboratory, California
  Institute of Technology, under contract with the National
  Aeronautics and Space Administration.
\end{acknowledgments}


\begin{thebibliography}{}
  
\bibitem[Armstrong, Estabrook, \& Tinto, 1999] {AET} J.W. Armstrong,
  F.B. Estabrook, \& M. Tinto, {\it ApJ}, {\bf 527}, 814 (1999).
  
\bibitem[Estabrook, Tinto, \& Armstrong, 2000] {ETA} F.B. Estabrook,
  M. Tinto, \& J.W. Armstrong, {\it Phys. Rev. D}, {\bf 62}, 042002
  (2000).
  
\bibitem[Tinto, Estabrook, \& Armstrong, 2002] {TEA} M. Tinto, F.B.
  Estabrook, \& J.W. Armstrong, {\it Phys. Rev. D}, {\bf 65}, 082003
  (2002).
  
\bibitem[Tinto, Armstrong, \& Estabrook, 2000] {TAE} M. Tinto, J.W.
  Armstrong, \& F.B. Estabrook {\it Phys. Rev. D}, {\bf 63}, 021101
  (R) (2000).
  
\bibitem[Dhurandhar, Nayak, \& Vinet, 2002] {DNV} S.V. Dhurandhar,
  K.R. Nayak, \& J.-Y. Vinet {\it Phys. Rev. D}, {\bf 65}, 102002
  (2002).
  
\bibitem[Pre-Phase A Report, 1998] {PPA98} P. Bender, K. Danzmann, \&
  the LISA Study Team, {\it{Laser \ Interferometer \ Space \ Antenna \ 
      for \ the \ Detection \ of \ Gravitational \ Waves, \ Pre-Phase
      \ A \ Report}}, $\bf{MPQ 233}$ (Max-Planck-Instit\"ut f\"ur
  Quantenoptik, Garching), (1998). 

\bibitem[Noble, 1969] {Noble69} B. Noble, {\it  Applied Linear Algebra},
  Prentice/Hall International, p. 378, (1969).
  
\bibitem[Selby, 1964] {Selby64} S. Selby, {\it Standard of
  Mathematical Tables}, The Chemical Bubber Co., p. 131, (1964). 

\bibitem[Finn, 2001] {Finn01} L.S. Finn, {\it Phys. Rev. D}, {\bf 63},
  102001, (2001).
  
\bibitem[Wolf, 2002] {Wolf02} S. Wolfram, {\it Mathematica: User
    manual}, Wolfram Research, Inc., (2002).
  
\bibitem[Tinto, Estabrook, \& Armstrong 2002] {WP02} M. Tinto, F.B.
  Estabrook, \& J.W. Armstrong, {\it LISA Pre-Project Publication},
  $(http://www.srl.caltech.edu/lisa/tdi\_wp/LISA\_Whitepaper.pdf)$,
  (2002).

\end{thebibliography}
\end{document}